\documentclass[12pt,a4paper]{article}
\usepackage{epsfig}
\usepackage{afterpage}
\usepackage{float}
\restylefloat{figure}
\begin{document}
\newcommand{\bea}{\begin{eqnarray}}
\newcommand{\eea}{\end{eqnarray}}
\renewcommand{\thefootnote}{\fnsymbol{footnote}}
\newcommand{\grgl}{\:\hbox to -0.2pt{\lower2.5pt\hbox{$\sim$}\hss}
{\raise3pt\hbox{$>$}}\:}\renewcommand{\thefootnote}{\fnsymbol{footnote}}
\newcommand{\klgl}{\:\hbox to -0.2pt{\lower2.5pt\hbox{$\sim$}\hss}
{\raise3pt\hbox{$<$}}\:}\renewcommand{\thefootnote}{\fnsymbol{footnote}}
%
\newsavebox{\tempbox}
\def\mycaption#1{
\sbox{\tempbox}{#1}
\vskip0.5cm
\ifdim \wd\tempbox >\hsize
#1
\else
\centering #1  
\fi
}
\def\tabcaption#1{
\par
{\centering\parbox{12cm}{\refstepcounter{table}
\mycaption{\footnotesize{\tablename\ \thetable: #1}}}
}}
\def\figcaption#1{
\par
\hspace*{8mm}{\centering\parbox{12cm}{\refstepcounter{figure}
\mycaption{\footnotesize{\figurename\ \thefigure: #1}}}
}}
\def\myfigcaption#1{
\par
\vspace*{-0.1cm}
\hspace*{8mm}{\centering\parbox{12cm}{\refstepcounter{figure}
\mycaption{\footnotesize{\figurename\ \thefigure: #1}}}
}}
\bigskip
\vspace*{1.5cm}\begin{center}
{\large\bf{Electroweak Phase Transition in the Early Universe ?}}
\end{center}
\bigskip
\vspace*{1cm}\begin{center}
 Bastian Bergerhoff\footnote{e-mail: B.Bergerhoff@thphys.uni-heidelberg.de; Supported by the DFG, Address after Nov. 1$^{\mathrm{st}}$, 1997: Physik-Department, TU-M{\"u}nchen, D-85748 Garching} and Christof Wetterich\footnote{e-mail: C.Wetterich@thphys.uni-heidelberg.de}\\
 Institut f\"ur Theoretische Physik \\
 Universit\"at Heidelberg \\
 Philosophenweg 16, D-69120 Heidelberg
\end{center}
\setcounter{footnote}{0}
\bigskip
\vspace*{1cm}\begin{abstract}
\noindent
Existence and properties of the electroweak phase transition
in the early universe depend strongly on the mass of the
Higgs scalar $M_H$. There is presumably no true symmetry restoration
at high temperature. 
Nevertheless, a first order phase transition occurs in the standard model for $M_H \klgl 70$~GeV.
For a realistic scalar mass $M_H \grgl 70$~GeV the transition to the high temperature
regime is described by a crossover, due to the strong electroweak gauge interactions for temperatures near and above the critical temperature.
Electroweak baryogenesis
during this transition seems not possible within the standard
model. 
The observed baryon asymmetry in the universe therefore implies the necessity of an extension of the standard model.
\end{abstract}
\vspace*{1cm}\newpage
During the early stages of the evolution of the universe the matter was in a hot plasma state, with density and temperature connected by $\rho \propto T^4$.
Within the standard hot big bang cosmology the time evolution is $\rho \propto t^{-2}$ and the universe must once have been in a state where $\rho$ was much larger than nuclear density.
A description in terms of hadrons must break down for such high densities and one expects matter to be in a new phase (currently called the quark-gluon-phase).
At even earlier times $t \ll 10^{-12}$~sec the temperature also exceeded the Fermi scale $<\varphi> \simeq 175$~GeV characteristic for spontaneous symmetry breaking in the electroweak sector of the standard model.
In close analogy to many statistical systems it has been speculated that at very high temperatures the spontaneously broken $SU(2)$-symmetry gets restored 
\cite{KirzhnitsLinde}.
This would reflect the trend that at high temperatures a system exhibits less order and more symmetry.
In the simplest picture the expectation value of the Higgs doublet $\varphi$ vanishes at high $T$.
Restoration of $SU(2)$-symmetry in particular means that the $W^\pm$ and a linear combination of the $Z^0$ and the photon form a triplet with degenerate mass.
Similarly, the left handed bottom and top quarks become indistinguishable, forming a doublet, and this holds even for such different particles as the (left handed) electrons and neutrinos!
One may naively think that the bottom and top quarks should even be massless if the chiral $SU(2) \times U(1)$ symmetry forbidding mass terms for these particles is restored at high temperature.
The relevant excitations are, however, pseudoparticles which correspond to excitations of the high temperature plasma.
In such a thermal equilibrium state Lorentz-symmetry is not conserved - the heat bath singles out a rest frame - and the remaining space-time symmetries are three dimensional rotations plus translations.
These symmetries allow for a mass term for the quark- and lepton-pseudoparticles which is near $\pi T$ for the lowest excitations. 
Naively, one may also guess that at high $T$ the mass of the $W$ and $Z$-bosons vanishes since for $T = 0$ their mass is generated by spontaneous symmetry breaking $<\varphi> \neq 0$.
The way how the corresponding pseudoparticles get masses at large $T$ is more subtle and will be explained below.

The universe must have undergone at least two important qualitative transitions as it has cooled down in its early history from a very high temperature state (say $T = 1$ TeV) to a temperature of a few MeV corresponding to nucleosynthesis. 
During the first - the electroweak phase transition - $SU(2)$-symmetry was spontaneously broken by the Higgs mechanism, giving to the quarks, leptons and gauge bosons the masses observed in our environment.
The typical transition temperature should be of the order of the Fermi scale or $\simeq 100$~GeV, details depending on the mass of the Higgs particle.
The second of these transitions is related to the dynamics of the strong interactions and occured at $T \simeq 100$ MeV.
At this temperature the (approximate) chiral symmetry of QCD was spontaneously broken and the mesons and hadrons acquired the properties observed today.
Even though we will often use the word ``phase transition'' generically for a rapid qualitative change in the particle properties, it should be emphasized that it is by no means clear that these transitions must be phase transitions in the more strict sense.
A priori, we do not know if some quantities are discontinuous (as the order parameter at a first order transition) or some response functions diverge (as the correlation length or inverse mass at a second order transition).

Beyond the great conceptual interest of these cosmological phase transitions for particle physics - they involve the dynamics of spontaneous symmetry breaking which is a keystone in modern particle physics - there may also be very interesting cosmological consequences:
The reason is that such a transition may create an out of equilibrium situation!
For most of its evolution the big bang universe realizes a local thermal equilibrium.
This implies that the densities of all particles in thermal equilibrium are simply given by the corresponding Boltzmann factors.
The universe therefore has lost memory about most of the details of its state in earlier epochs.
This makes cosmology predictive, but it also severely limits our capacity to learn from cosmological observations many details of the particle physics which has governed the universe in very early epochs.
Only a few quantities escape from this rule.
A prominent example, to which we owe our existence, is the asymmetry $\Delta B$ between the densities of baryons and antibaryons.
The difference between the number of baryons and antibaryons remains conserved for all times sufficiently after the electroweak phase transition since the baryon number violating interactions in the standard model of particle physics are simply too weak
\cite{tHooft}
in order to enforce the thermal equilibrium value $\Delta B = 0$.
The physics that leads to a value $\Delta B > 0$ after the electroweak phase transition can therefore be tested by observing todays excess of matter over antimatter.
On the other hand, for temperatures above the critical temperature of the electroweak phase transition the rate of baryon number violating processes was sufficient
\cite{KuzminRubakovShaposhnikov}
for them to be in thermal equilibrium.
As a consequence, the value of $\Delta B$ before the electroweak phase transition can be predicted as a function of the asymmetry in baryon minus lepton number $\Delta (B-L)$.
The latter is conserved by all interactions of the standard model (except for a possible very weak violation due to nonvanishing neutrino masses which is not relevant in our context).
In particular, for $\Delta (B-L) = 0$ one has $\Delta B = 0$.
We are therefore left with two alternative scenarios:
Either the cosmology at times much earlier than the electroweak phase transition, when interactions beyond those contained in the standard model played a role, has produced a nonvanishing asymmetry $\Delta (B-L) \neq 0$.
This can lead to today's $\Delta B > 0$ even without ever leaving thermodynamic equilibrium at the electroweak phase transition.
Or else $\Delta (B-L) = 0$, and in this case a nonvanishing baryon asymmetry has to be produced during (or after) the electroweak phase transition.
Clearly, creating the baryon asymmetry during the electroweak phase transition (``electroweak baryogenesis'')
\cite{KuzminRubakovShaposhnikov,ShaposhnikovDeltaB}
needs an out of equilibrium situation
\cite{Sakharov}.

It is at this point that the order of the electroweak phase transition becomes of crucial importance:
For a first order transition the (coarse grained) free energy has near the critical temperature $T_c$ two separated local minima.
For the electroweak transition they are distinguished by different values of the scalar field $\varphi$.
For $T > T_c$ the lowest minimum corresponds to the high temperature phase.
(Pseudo)particle properties are determined by an expansion around this minimum and may be quite different from what we are used to -- see the case of symmetry restauration discussed above.
As the universe cools below $T_c$ the lowest minimum jumps to the one of the low temperature phase.
This may be associated with ``our vacuum'' since the properties of the excitations around this minimum determine at $T=0$ the observed constants of particle physics.
A barrier\footnote{The barrier often depends on details of the coarse graining, for details see 
\cite{BergesTetradisWetterich}.
}
between the two minima forbids, however, a simple smooth transition at $T = T_c$.
Typically, the transition occurs through the formation of droplets (or ``bubbles'') of our vacuum, very similar to the condensation of vapor.
These bubbles expand almost with the speed of light, scatter on each other and melt until the whole universe is filled with the vacuum corresponding to the low temperature phase.
Such a first order transition is a short dramatic period in the evolution of the universe, with many processes out of equilibrium.
A second order transition is much smoother.
There is always only one minimum of the coarse grained free energy and order parameters or thermodynamic quantities are continuous during the transition.
The phase transition is signaled by an infinite correlation length.
Also the temperature dependence of the order parameter, some particle masses and couplings or the specific heat are not analytical for $T \rightarrow T_c$.
No deviation from local thermodynamic equilibrium is expected.
Finally, we should also consider the possibility that the transition is no true phase transition at all, but rather an ``analytical crossover''.
In a crossover situation many quantities change rapidly in the transition region, but everything remains analytical and correlation lengths are finite.
Local thermodynamic equilibrium is again realized.
In conclusion, the idea of electroweak baryogenesis requires as a fundamental condition that the electroweak phase transition be sufficiently strong first order.

For a determination of the order of the electroweak phase transition we need to compute the (coarse grained) free energy or, equivalently, the temperature dependent effective potential $U(\rho,T)$ as a function of a constant Higgs field $\varphi$ ($\rho = \varphi^\dagger \varphi$) and temperature.
For $T=0$ this potential is well approximated by
\bea
U(\rho,0) = \frac{1}{2} \lambda (\rho-\rho_0)^2 \qquad , \qquad \lambda = \frac{M_H^2}{2 \rho_0}
\label{1}
\eea
with $\rho_0 = (175$~GeV$)^2$ and $M_H$ the mass of the Higgs scalar.
The value of $M_H$ will turn out to be of crucial importance for the characteristics of the electroweak phase transition.
At present we only have a lower experimental bound of $M_H > 70$~GeV 
\cite{Jerusalem}
in the standard model.

For a computation of the temperature dependence of $U$ one may first use perturbation theory.
The dominant effect is the generation of a term linear in $\rho \propto T^2$, 
\bea
\Delta U_1 = \frac{1}{16} \left( 3 g^2 + 4 \lambda + 4 h_t^2 \right) T^2 \rho
\label{1a}
\eea
such that the mass term at the origin becomes
\bea
U'(0,T) \equiv \frac{\partial U}{\partial \rho}(0,T) = -\lambda \rho_0 + \frac{1}{16} \left( 3 g^2 + 4 \lambda + 4 h_t^2 \right) T^2
\label{2}
\eea
Here $g$ is the electroweak gauge coupling, $h_t$ the Yukawa coupling of the top quark and the Yukawa couplings of the lighter quarks as well as electromagnetic effects are neglected.
A potential of the type (\ref{1}),(\ref{1a}) describes a second order phase transition, with $T_c$ determined by $U'(0,T_c) = 0$.
This gives already a quite good estimate of the critical temperature.
The next to leading correction in perturbation theory is nonanalytical in $\rho$,
\bea
\Delta U_2 = - \frac{3\sqrt{2}}{16\pi} g^3 T \rho^{3/2} + ...
\label{3}
\eea
with the dots standing for omitted contributions from scalar fluctuations.
Studying the shape of the combined potential (\ref{1}), (\ref{1a}) and (\ref{3}) for different values of $T$ it is easy to see that the nonanalytic term $\propto \rho^{3/2}$ leads now to a first order phase transition!
The discontinuity is, however, not very strong, with a tendency to weaken for larger $M_H$.
In this order of perturbation theory the discontinuity would be sufficient for electroweak baryogenesis only for $M_H$ substantially below the present experimental bound
\cite{ShaposhnikovRubakov}.

The validity of perturbation theory may, however, be questioned.
Defining a field dependent quartic scalar coupling $\lambda(\rho) = U''(\rho)$, we see that this quantity diverges for $\rho \rightarrow 0$ due to the nonanalyticity of (\ref{3}).
This coupling should determine the effective four scalar interactions in the high temperature phase if the minimum of $U(\rho,T)$ is at $\rho=0$.
Another way to identify the problems is to look at the effective expansion parameters in the perturbative series.
Due to the infrared structure of the loop expansion, they are $g^2 T / m_R(T)$ or $\lambda T / m_R(T)$ with $m_R(T)$ some relevant temperature dependent particle mass.
For $T$ approaching $T_c$ from below, the scalar mass can become very small.
For the high temperature phase the perturbative (magnetic) $W$-boson mass even vanishes at $\rho=0$.
Already early investigations
\cite{Linde,GrossPisarskiYaffe,BuchmuellerEtAl, ZwirnerEtAl}
have revealed that the perturbative expansion becomes uncontrolled in the high temperature phase and near $T_c$ in the low temperature phase if $M_H$ is larger than some value of the order of $70$~GeV.
These problems are clearly related to the infrared behaviour of the model in situations where some of the masses are small.
On the other hand, the low temperature phase should be well described by perturbation theory if $M_H$ is small enough, even in the vicinity of the critical temperature.

By now, several groups have performed perturbative calculations in two loop order 
\cite{2Loop}.
Most of these calculations are performed in the pure $SU(2)$-Higgs model, i.e.~with vanishing Yukawa couplings and electromagnetic coupling.
The value of the scalar mass $\overline{M}_H$ in this model is related to the scalar mass $M_H$ in the standard model by a perturbative calculation.
Whereas a strict Taylor expansion in the couplings $g$ and $\lambda$ yields only poor results, appropriate resummation schemes have led to a much better convergence and to a convincing agreement with the results from lattice simulations
\cite{lattice}
whenever perturbation theory is supposed to be valid.
Moreover, there often is a surprising agreement of some quantities with the simulations even for $\overline{M}_H$ as large as $60 - 70$~GeV and $T$ near $T_c$ where a fast convergence does not seem guaranteed a priori.
Of course, the systematic character of perturbation theory is absent in the resummation schemes.
Different such schemes give answers that differ at higher loop orders.
We will see below that physical quantities are always analytic in the couplings $g$, $\lambda$ and $h_t$ (except for a special critical value $M_H^{(c)}$ corresponding to the ``endpoint'' of a line of first order transitions).
This makes statements about the size of ``higher order'' or ``nonperturbative'' effects depending on the loop order in perturbation theory and, furthermore, on the resummation scheme.
The higher order or nonperturbative effects are always large if a strict Taylor expansion in the couplings is used, whereas they can be made, in principle, arbitrarily small within an ``optimal resummation scheme''.
The convergence of the perturbative series depends strongly on which thermodynamic quantity is computed.
The most robust quantity is the critical temperature $T_c$ which can typically be found within a few percent accuracy.
The convergence is also very satisfactory for the jump in the order parameter or the related latent heat provided $\overline{M}_H < 70$~GeV.
Here the accuracy diminishes with increasing $\overline{M}_H$ and the perturbative series shows only a slow or no convergence for $\overline{M}_H$ around $70$~GeV or larger.
Finally, it is very hard to compute the surface  or interface tension $\sigma$ (the energy per area at a boundary between the two phases) reliably in perturbation theory.
This is important, since the surface tension governs the dynamics of bubbles in a first order phase transition.
We will understand later why some quantities are much more sensitive to ``nonperturbative'' effects than others.

If the fermions are neglected, the high temperature standard model can be simulated by lattice Monte-Carlo methods without major conceptual problems.
In the vicinity of $T_c$ the fermions play no role in the long distance dynamics.
Their influence on the value of $T_c$ can be computed perturbatively.
Access to situations with a large correlation length (for $T$ near $T_c$ and $\overline{M}_H$ near $70$ GeV) remains nevertheless difficult in direct simulations.
(Finite size techniques for a series of lattices with different physical volume could be employed.)
A major effort of several groups
\cite{lattice}
to simulate the $SU(2)$-Yang-Mills theory with scalar fields (fermions and electrodynamics are neglected) has produced rather accurate results for several thermodynamic quantities relevant for the electroweak phase transition. 
\bigskip
\\
\hspace*{3cm}{\centering{
\begin{tabular}
{|c|c|c|c|} 
\hline
 $\overline{M}_H$ [GeV]  & 35 & 60 & 70 \\
\hline
\hline
 $\left.T_c\right|_{\rm latt}$ [GeV] & 92.64(7) & 138.38(5) & 154.52(10) \\
 $\left.T_c\right|_{\rm pert}$ [GeV] & 93.3 & 140.3 & 157.2 \\
\hline
 $\left.(v/T_c)\right|_{\rm latt}$ & 1.86(3) & 0.674(8) & 0.57(2) \\
 $\left.(v/T_c)\right|_{\rm pert}$ & 1.87 & 0.82 & 0.70 \\
\hline
 $\left.(\sigma/(T_c)^3)\right|_{\rm latt}$ & 0.0917(25) & 0.0023(5) &  \\
 $\left.(\sigma/(T_c)^3)\right|_{\rm pert}$ & 0.066 & 0.0078 & 0.0049 \\
\hline
\hline
\hline
\end{tabular}}}\\
\bigskip

\noindent A comparison
\cite{KajantieEtAlComparison}
between lattice results and a particular resummed two loop perturbative calculation is quoted in the table, with $v \simeq \sqrt{2 \rho_0(T_c)}$ the jump in the order parameter\footnote{More precisely, $v$ is a gauge invariant quantity which is related to the doublet expectation value in a specific gauge (i.e.~$\rho_0$) by perturbation theory.}.
The errors in the lattice simulations do not include systematic shifts in $T_c$ and $M_H$ which depend on the precise top quark mass and are, more generally, model dependent in extensions of the standard model\footnote{The numbers given here are for $h_t = 0$ with electromagnetic effects neglected. They are related to physical parameters in a given theory by corrections that are well controlled in perturbation theory
\cite{KajantieEtAlComparison}. For the standard model with $m_t = 175$ GeV the values of the pole mass of the Higgs at vanishing temperature, $M_H$,  corresponding to $\overline{M}_H$ given above are $M_H = 68.0(5)$ GeV for $\overline{M}_H = 70$ GeV and $M_H = 51.2(5)$ GeV for $\overline{M}_H = 60$ GeV. The errors are due to dimensional reduction, vacuum renormalization and a perturbative estimate of the effect of the extra $U(1)$. In the standard model there is no physical value corresponding to $\overline{M}_H = 35$ GeV. For the standard model ($h_t \neq 0$) the temperature also has to be rescaled. Thus e.g.~the critical temperature for $\overline{M}_H = 70$~GeV corresponds to $T_c^{(\mathrm{SM})} = 104.8$~GeV.}.
Also the effects of long distance photon fluctuations are omitted.
Both lattice and perturbative computations quoted in the table have actually been performed for an effective three dimensional $SU(2)$-Yang-Mills type model (see below) which can be connected to the standard model at large temperatures.
Similar comparisons can be found in 
\cite{JansenComparison}.

Let us now turn to the question why an understanding of the electroweak phase transition needs nonperturbative methods despite the smallness of the couplings $g$, $\lambda$ and $h_t$.
The first key observation notices that the physics of fluctuations with momenta $|p| \ll \pi T$ is governed effectively by classical statistics.
The effects of quantum statistics are small for these modes and, correspondingly, their interactions can be well described by three dimensional field theories.
It is known since a long time 
\cite{AppelquistNadkarniDimRed}
that for high enough temperature one can reformulate equilibrium quantum field theory as a three dimensional effective field theory (``dimensional reduction'').
The quantitative relevance of this effect for the electroweak phase transition and the connection with the shortcomings of perturbation theory were realized, however, only a couple of years ago
\cite{TetradisWetterich,ReuterWetterichNP,PatkosEtAl,KajantieEtAlNP}.
In fact, quantum field theory in thermal equilibrium can be described by a Euclidean field theory with time on a torus of circumference $1/T$.
The (Euclidean) zero-components of the momenta are therefore discrete, $p_0 = 2 \pi n T$, with integers $n$, and are called Matsubara frequencies\footnote{For fermions $n$ should be replaced by $n+\frac{1}{2}$ due to the antisymmetric boundary conditions. The lowest value of $p_0$ is therefore $\pi T$. This gives rise to a temperature dependent effective mass for the lowest excitation and explains why fermions usually play no role in the effective three dimensional theory.}.
It is very intuitive that for length scales much larger than $1/(\pi T)$ or momenta much smaller than $\pi T$ one cannot resolve the time dimension anymore and is left with an effective three dimensional description.

There are different ways of performing dimensional reduction to obtain an effective three dimensional picture.
One possibility is to integrate out quantum and thermal fluctuations with momenta $p_0^2 + \vec{p}^{\,2}$ larger than some infrared scale $k^2$ and compute first the coarse grained free energy with $k_T \sim \pi T$.
If the modes with $p_0^2 + \vec{p}^{\,2} < k_T^2$ are included next, i.e.~if $k$ is subsequently lowered from $k_T$ to zero, one sees that the change of the effective couplings is now dominated by contributions from the lowest Matsubara frequency $n=0$.
The running of the couplings becomes three dimensional.
This procedure was first used in the electroweak context 
\cite{TetradisWetterich,ReuterWetterichNP,Tetradis}
and is very useful for the study of flow equations for running couplings.
For example, if the gauge bosons are neglected, the resulting three dimensional theory is in the universality class of the $O(4)$ Heisenberg model.
The phase transition is second order and the associated critical exponents and critical equation of state are a consequence of the effective three dimensional running coupling constants being attracted to the Wilson-Fisher fixed point.
In the approach described above, this can be seen clearly in the context of high temperature field theory
\cite{TetradisWetterich}.
The apparent infrared divergences in a perturbative treatment of this model for $T \rightarrow T_c$ are now cured by the fact that the running effective renormalized quartic scalar coupling $\lambda_R$ actually approaches zero for $T \rightarrow T_c$, with $\lambda_R(T) T / m_R(T)$ going to a constant as the inverse correlation length $m_R$ vanishes.
(Perturbation theory is an expansion in $\lambda T / m_R(T)$ with $\lambda$ a fixed coupling, such that $\lambda T / m_R$ diverges for $m_R(T\rightarrow T_c) \rightarrow 0$.)
For $g>0$ the phase transition will generically not be of second order - except for a critical scalar mass $M_H^{(c)} \simeq 70$ GeV (see below) for which an effective scalar model becomes relevant in the immediate vicinity of $T_c$.

Another method of dimensional reduction, more suitable for perturbation theory and effective three dimensional lattice simulations, consists in integrating out all Matsubara frequencies with $n \neq 0$.
This step can be done within the validity of perturbation theory, for example in a one loop calculation.
The resulting three dimensional theory describes the dynamics of the $n=0$ modes.
At this stage it is still ultraviolet divergent, but the renormalized couplings of the three dimensional theory can be mapped onto the ones of the high temperature field quantum field theory by computing suitable (infrared safe) physical quantities in both the effective and the full theory.
This method is developed by now to a high level of sophistication
\cite{PertDimensionalReduction}.
We emphasize that for both ways of dimensional reduction the difficult infrared behaviour of the model is now described within a simplified three dimensional model.
For example, the three dimensional model does not contain quarks and leptons anymore and also the $W_0$-component of the gauge fields $W_\mu$ can be integrated out.
The separation of the physics at different length scales and the simplified description of the problematic long distance part has proved crucial for a quantitative understanding of the electroweak phase transition.
What remains at this stage is a solution of the three dimensional field theory.

This brings us to the second key element for the qualitative understanding of the characteristics of the electroweak phase transition:
In the high temperature phase, and also in the low temperature phase for $T$ near $T_c$ and $M_H \grgl 70$~GeV, the dynamics is dominated by effective {\it{strong}} interactions
\cite{ReuterWetterichNP,BergerhoffWetterichNP}!
This statement may surprise at first sight, since all couplings of the underlying electroweak standard model are weak.
Nevertheless, we should look at the strength of temperature dependent renormalized couplings $g_R(k,T)$ and $\lambda_R(k,T)$, as defined, for example, by the interactions of three gauge bosons or four scalars in thermodynamic equilibrium at temperature $T$, with an effective infrared cutoff $k$ in the vertex given either by external momenta or the mass of the involved particles.
Typically, $g$ and $\lambda$ (we omit the subscript $R$ in the following) will only depend on $k/T$ if the slow logarithmic running of the $T=0$ (four dimensional) couplings is neglected.
The pure three dimensional nonabelian gauge theory (without the Higgs particle) is a confining theory, with a running gauge coupling given for $k \klgl \pi T$ by
\cite{ReuterWetterichBeta}
\bea
k\frac{\partial}{\partial k} g^2(k,T) = - \frac{23 \tau}{24 \pi} g^4(k,T) \frac{T}{k} - ...
\label{4}
\eea
Here $\tau$ is a constant of order one which depends on the particular definition of the renormalized gauge coupling or the scale $k$.
What is new as compared to the well known four dimensional running is the factor $T/k$ which reflects the different infrared behaviour of the three dimensional loop diagrams.
As a result, the $k$-dependence of $g$ follows a power-like behaviour rather than the logarithmic behaviour for $T=0$.
We can start at $k_T = \pi T$ with the zero temperature value $g^2(\pi T,T) \sim g^2(\pi T,0) = g_4^2 \simeq \frac{4}{9}$ since temperature plays no major role for the fluctuations with $p^2> (\pi T)^2$.\vspace*{0.5cm}\\
\hspace*{1.3cm}\epsfig{file=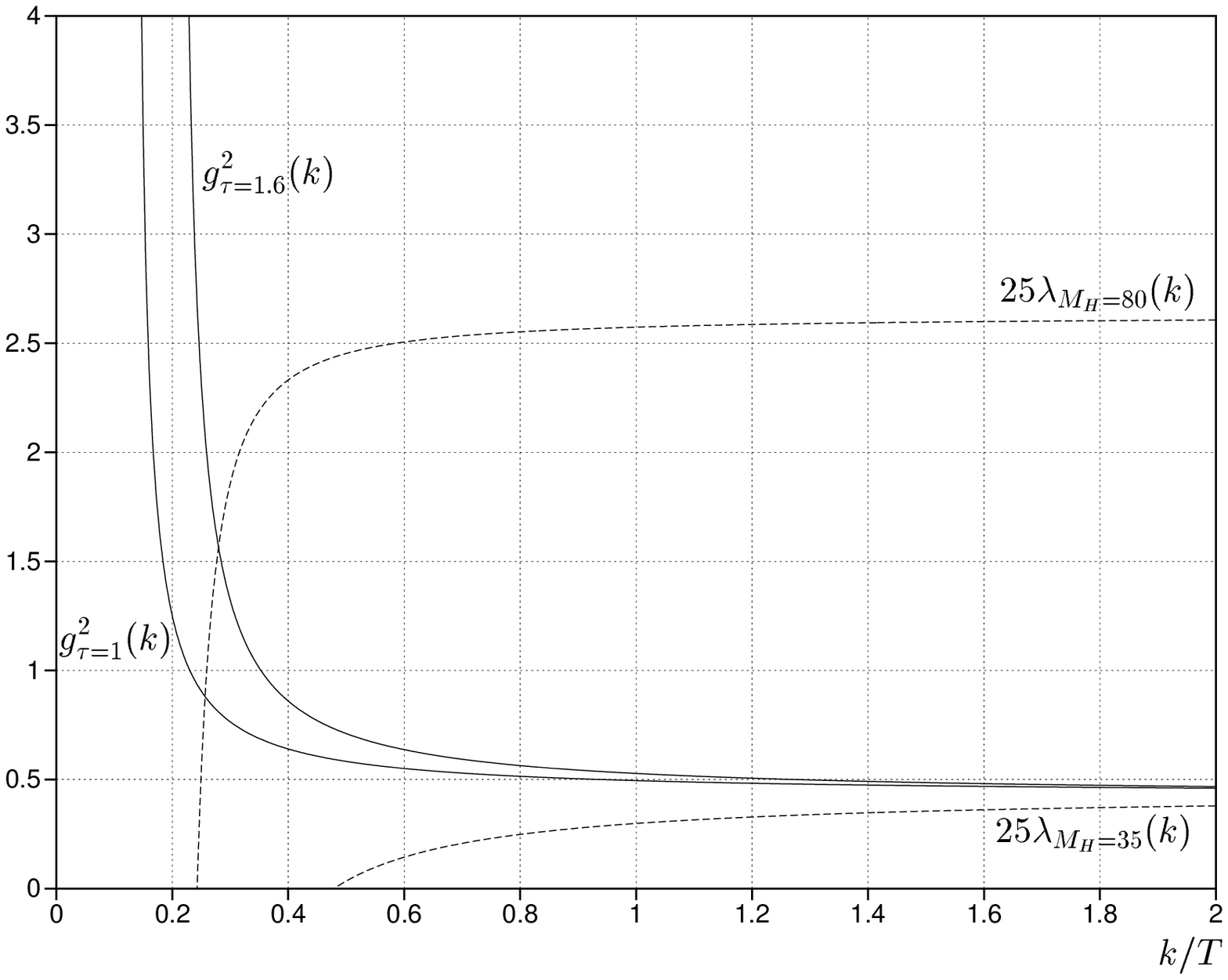,width=11.5cm}
\figcaption{The running gauge coupling $g^2(k)$ for two different values of $\tau$ (see below) (solid lines) and the running scalar coupling $\lambda(k)$ for $\overline{M}_H = 35$ GeV and $\overline{M}_H = 80$ GeV (broken lines).}\vspace*{1cm}\\
In figure 1 we have plotted the $k$-dependence of $g^2$ (solid lines), together with the running of the quartic scalar coupling $\lambda$ for two initial values $\lambda(\pi T,T) \simeq \lambda (\pi T,0)$ corresponding to $\overline{M}_H = 35$~GeV and $80$~GeV (broken lines).
We see  that at 
\bea
k_s \simeq \frac{1}{2} g_4^2 T = 2\pi \alpha_w T \simeq 0.2 \,\,T
\label{5}
\eea
the effective gauge coupling becomes strong!
A scale in this order of magnitude may be associated with a three dimensional ``confinement scale'' in analogy to QCD.
Indeed, one may introduce a dimensionless three dimensional gauge coupling $g_3^2(k) = g^2(k,T) T / k$ and specify the model by the value of $g_3^2$ at some scale $k$, say $k=T$.
The scale of the theory is then set by the running of $g_3^2(k)$ just like in QCD.
One expects (at least generically) no massless particles in such a theory, with a typical mass of the lowest excitations $\sim k_s$.
This is the ``nonperturbative'' mechanism which provides a ``magnetic mass'' to the (perturbatively massless) $W$-bosons, $M_W \propto k_s \propto \alpha_w T$
\cite{ReuterWetterichNP,DoschKripfganzLaserSchmidt} .
The nonabelian nature of the gauge theory which leads to a ``confinement scale'' is crucial here.
If $g_3^2(k)$ would not run, for example due to an infrared fixed point as present in some abelian gauge theories
\cite{BergerhoffEtAlLargeNAHM}
(or analogously for the pure scalar theory),
the fact that $g_4^2 T$ has dimension of mass does {\it{not}} 
imply the existence of a physical mass scale.
We emphasize that the way how the perturbative infrared divergences are cured in a nonabelian Yang-Mills theory is quite different from the pure scalar theory for $T=T_c$.
In the Yang-Mills theory strong interactions induce a mass gap ($M_W \sim k_s$) which in turn stops the increase of $g_3^2$ even if the external momenta in the vertex go to zero.
In the scalar theory there is no mass gap and the increase of $\lambda$ for external momenta going to zero is stopped by an infrared fixed point.
Since $k_s$ is proportional to $\alpha_w$ the ``nonperturbative mass generation'' can actually also be caught within suitably adapted versions of resummed perturbation theory, as for example the solution of gap equations 
\cite{BuchmuellerPhilipsen}.
We also note that $k_s$ sets the scale for possible nonperturbative condensates, again in close analogy to QCD.

Once the scalar field is included, one still expects in the high temperature phase a strongly interacting gauge theory, with only minor modifications from the scalar field.
Since the effective number of degrees of freedom of the gauge fields is much higher than the one of the Higgs field, it was proposed 
\cite{WetterichSintra}
that the high temperature phase of the electroweak theory is very similar to the pure Yang-Mills theory (except for the existence of additional excitations involving the scalar field). For the low temperature phase the effective infrared cutoff is set by the (perturbative) $W$-boson mass,
\bea
k_W^2 = M_W^2(T) = \frac{1}{2} g^2(k_W,T) Z_\varphi(k_W,T) \rho_0(T)
\label{6}
\eea
with $\rho_0(T)$ denoting the minimum of $U(\rho,T)$ and $Z_\varphi$ the wave function renormalization constant for the scalar field.
As long as $k_W \gg k_s$, the gauge coupling is not strong in the low temperature phase.
This situation is realized for $M_H \klgl 70$~GeV, and this explains why quantities of the low temperature phase like $\rho_0(T)$ can be estimated reliably by perturbation theory in this case.
For small Higgs masses, also the running of the couplings as depicted in figure 1 does not have a major effect on quantities defined in the broken phase.
In figure 2 we display the results of a calculation of the temperature dependence of the vacuum expectation value of the Higgs field, $v(T) = \sqrt{2 \rho_0(T)}$, taking into account the running of the effective couplings by the solution of non-perturbative flow equations
\cite{BergerhoffWetterichNP}.
We also compare the results to the perturbative 1- and 2-loop results as given in 
\cite{2Loop}.
By including the running of the couplings, a partial summation of contributions to all orders in ordinary perturbation theory is performed.
For a detailed comparison one should note, however, that the effects of the running couplings do not account for all two-loop contributions.
\begin{figure}[H]
\hspace*{1.3cm}\epsfig{file=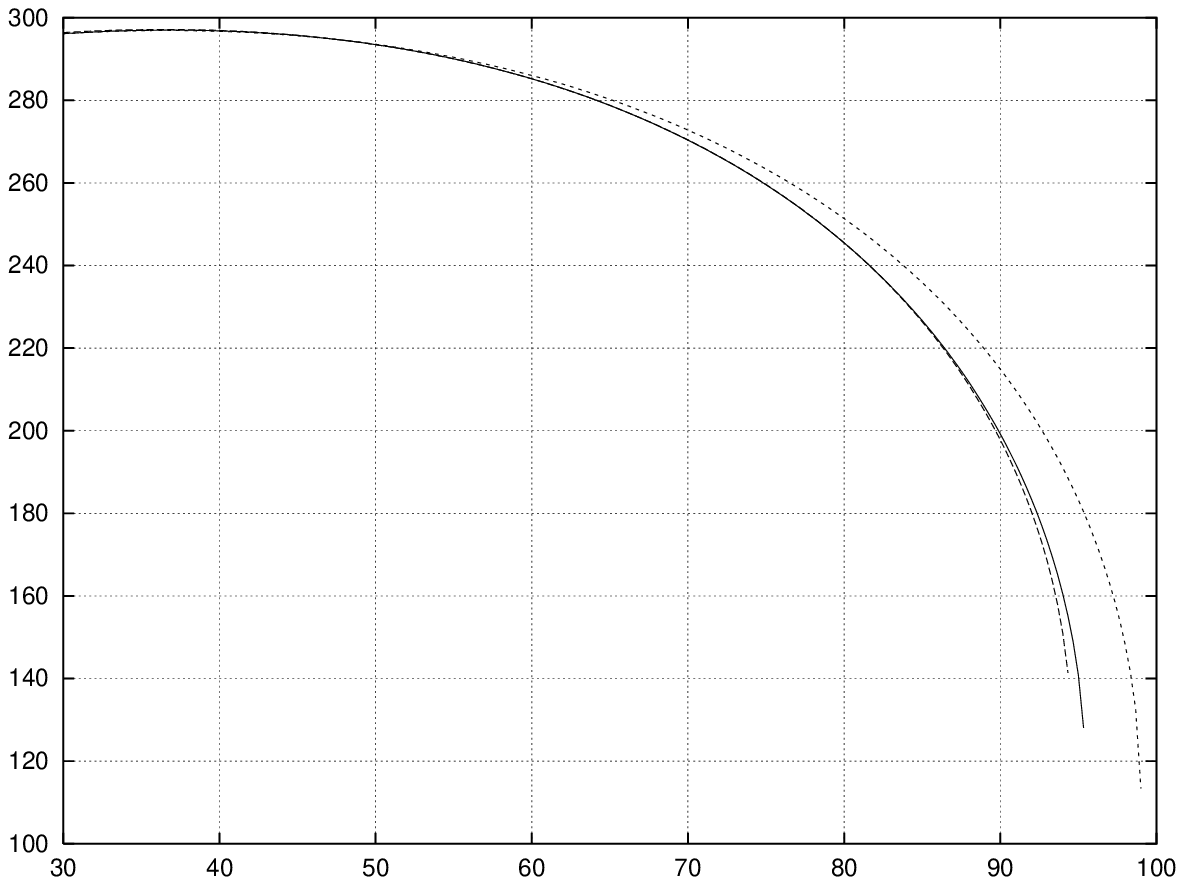,width=11.5cm}
\put(-20,-10){$\scriptstyle{T [\mathrm{GeV}]}$}
\put(-354,215){$\scriptstyle{v(T) [\mathrm{GeV}]}$}
\put(-50,140){$\scriptstyle{\mathrm{1-Loop}}$}
\put(-71,70){$\scriptstyle{\mathrm{2-Loop}}$}
\put(-57,45){$\scriptstyle{\mathrm{ERGE}}$}
\figcaption{ Scalar expectation value $v(T)$ for $\overline{M}_H = 35$ GeV from 1- and 2-loop perturbation theory and including the effects of the running of couplings (ERGE).}
\end{figure}
\noindent On the other hand, it was found
\cite{BergerhoffWetterichParis}
that for $M_H \grgl 70$~GeV and $T$ near $T_c$ the effective gauge coupling $g(k_W,T)$ has already grown to large values.
For large masses of the Higgs scalar one therefore predicts that strong interaction effects play a role both for the high and the low temperature phase.
The dynamics of the electroweak phase transition can then not be understood without understanding the strong interaction dynamics of the effective three dimensional theory!

How can this picture of a ``strongly interacting electroweak phase transition'' be reconciled with the partial success of perturbation theory (cf.~the table)?
At this point it is important to realize that strong three dimensional interactions are not in contradiction with an expansion in the small parameter $\alpha_w$.
This is different from what we are used to in QCD and can be traced back to the fact that the confinement scale itself is proportional to $\alpha_w$ (\ref{5}).
A typical nonperturbative contribution to the free energy at $\rho=0$ must by simple dimensional arguments be
\bea
\Delta U(\rho=0,T) \propto k_s^3 T \simeq \frac{1}{8} g_4^6 T^4
\label{7}
\eea
Correspondingly, one finds in a perturbative calculation of $U(0,T)$ that contributions up to $g^5$ are infrared finite. 
Only the four loop contribution $\propto g^6$ is infrared divergent - and the strong interaction dynamics exactly cures this infrared problem by providing a mechanism for the generation of a magnetic mass $\propto k_s$ for the $W$-bosons.
In this sense the strong interaction dynamics only fixes the otherwise undefined coefficients in an expansion of $U(0,T)$ in powers of $g^N$ for $N\geq 6$, but it remains fully consistent with a series expansion and no nonanalytic behaviour is introduced.
Even for a relatively large coefficient of the $\alpha^3 T^4$ contribution (note that there is no factor of $1/(4\pi)$ in (\ref{5})) this remains still a moderate correction as compared to typical one or two loop contributions $\propto T^4$ or $\alpha T^4$.
The critical temperature $T_c$ is determined by the requirement that the free energy in the high temperature phase equals the one in the low temperature phase.
We understand now why $T_c$ can be computed quite accurately in perturbation theory, since the ``nonperturbative'' uncertainty in the difference of the free energy between the two phases is again $\propto g_4^6 T^4$.

Another relation often associated with weak interactions or an ideal gas situation is the Stefan-Boltzmann law $\rho \propto T^4$ which is very important for cosmology.
One may wonder what happens to this relation when QCD or the weak interactions at high temperature are in fact described by strong effective three dimensional interactions.
The associated scale of strong interactions $k_s$ is, however, itself proportional to the temperature.
By simple dimensional arguments it is obvious that the law $\rho \propto T^4$ holds quite generally in the approximation where the temperature is the only relevant mass scale besides the Planck mass, i.e.~if particle masses are small as compared to $T$ and similar for typical scales associated with the running of couplings at $T=0$ as the confinement scale in QCD.
No assumption about the strength of couplings or an approximate ideal gas situation enters here.
Except for particle mass thresholds the dominant corrections arise from the (zero temperature) running of gauge and other couplings and are expected to be $\propto \alpha(T) T^4$ with $\alpha(T)$ the strong or weak fine structure constant $\alpha = g^2/(4\pi)$ evaluated at a momentum scale $T$.
For small gauge couplings this results in a logarithmic deviation $\rho = T^4 (c_1 + c_2 \alpha^2(\mu_0)\ln(T/\mu_0))$ in the temperature dependence of $\rho$ and it would be interesting to find out if this can be tested through some cosmological consequences.

We next turn to the jump of the ``order parameter'' $\rho_0$ between the low and high temperature phases.
For small enough $M_H$ we can use the perturbative value for $\rho_0$ in the low temperature phase and put $\rho_0=0$ in the high temperature phase.
Only for $M_H \grgl 70$~GeV the strong interaction dynamics influences substantially the low temperature phase.
This explains why the perturbative value of $v$ is reliable for low $M_H$ and also the increasing discrepancy between the perturbative and the lattice values in the second line in the table as $M_H$ increases.
Finally, the surface tension $\sigma$ is related to the $\rho$-dependence of $U(\rho,T)$ in the whole region between the minima relevant for the two phases.
The perturbative infrared divergences of $\partial U / \partial \rho$ occur already in two loop order $\propto g^4$ and there is no barrier in the lowest order potential (\ref{1}),({\ref{1a}).
This, together with the problems associated with the coarse graining in a strongly interacting effective theory
\cite{BergesTetradisWetterich},
explains the difficulty of a reliable perturbative computation of $\sigma$.
Quite generally, we note a ``hierarchy of robustness'' for perturbatively computed quantities which is related to the loop order at which perturbative infrared problems appear.
Quantities like the quartic scalar coupling ($\propto \partial^2 U / \partial \rho^2$) or the gauge coupling are already infrared divergent in one loop order and renormalization group methods should be used to deal with this problem.
In summary, the good partial agreement of perturbative results with lattice simulations is perfectly consistent with the picture of strong interactions at the electroweak phase transition.
These strong interactions are actually the ingredient needed to cure the partial shortcomings of perturbation theory.

The perhaps most spectacular consequence of the running coupling $g(k,T)$ becoming strong is the suggestion 
\cite{ReuterWetterichNP,WetterichSintra,BergerhoffWetterichNP}
that for large values of $M_H$ there is no genuine phase transition anymore even for (arbitrarily) small $\alpha_w$.
The high- and low temperature properties of the electroweak interactions are then connected by an analytic crossover.
This is based on an earlier observation
\cite{BanksRabinovici,FradkinShenker}
that there is no true order parameter for the electroweak phase transition since gauge symmetries are always conserved, only realized in different ways in situations of confinement or the Higgs phase associated with ``spontaneous symmetry breaking''.
There may therefore be an analytical connection between confinement and Higgs phase and it was argued that such a crossover is indeed realized in the three dimensional $SU(2)$-Higgs model if the gauge coupling is strong.
Indications in this direction were seen in early lattice simulations
\cite{DamgaardHeller,EvertzJersakKanaya}.
The general structure of the phase diagram of the $SU(2)$-Higgs model (with arbitrary $g_4^2$) was carefully discussed in 
\cite{EvertzJersakKanaya}
and we have drawn a projection for fixed $M_H$ (or $\rho_0$) and $\lambda$ in figure 3.\hspace*{0.5cm}\\
\hspace*{1.3cm}\epsfig{file=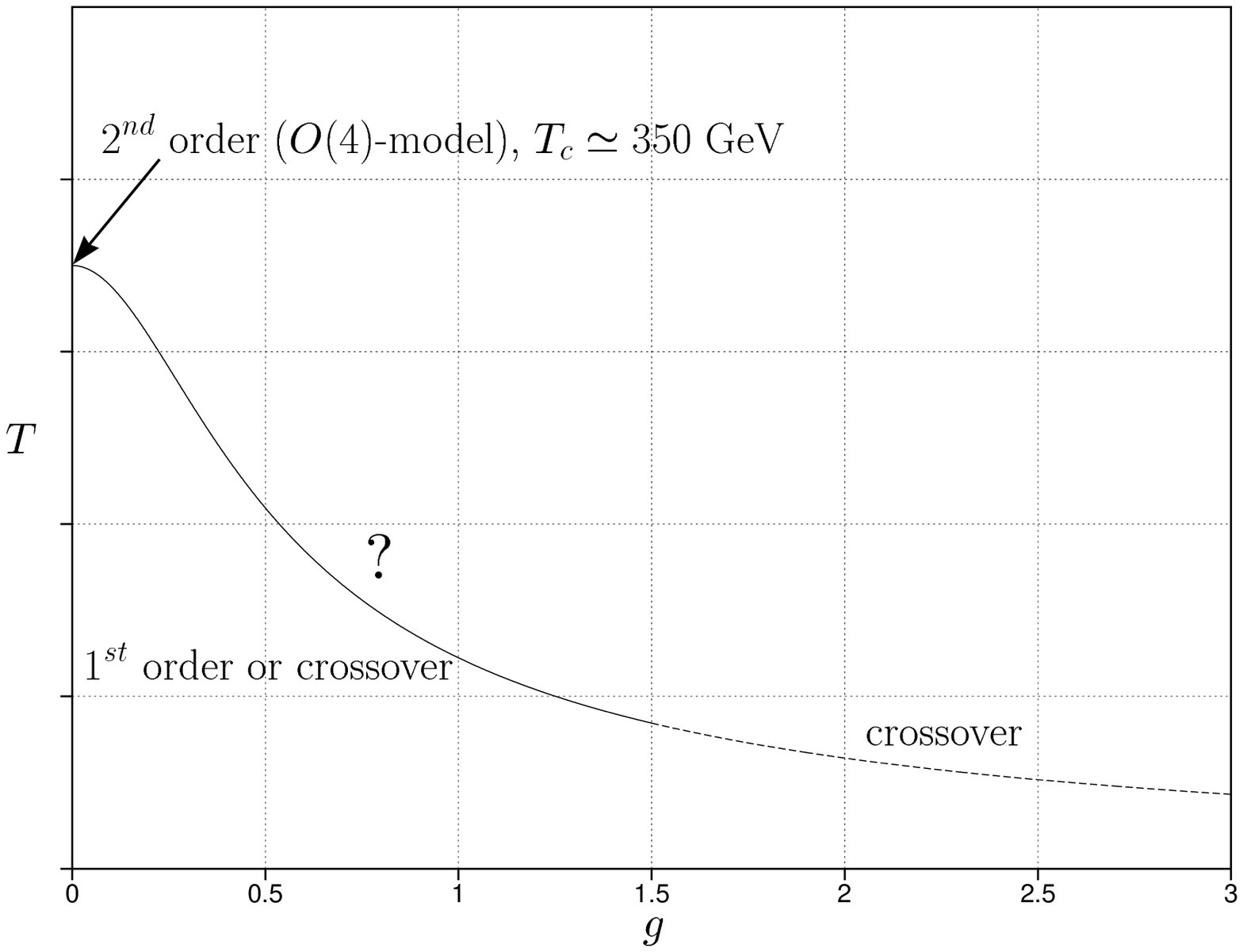,width=11.5cm}
\figcaption{The phase diagram for the $SU(2)$-Yang-Mills Higgs model. The transition curve is approximately $T_c^2 = 16 \lambda \rho_0 / \left( 3 g^2 + 4 \lambda + 4 h_t^2 \right)$ (see eq. (\ref{2})) and drawn here for $M_H = 80$ GeV, $h_t = 0$. The dashed line corresponds to the crossover observed in lattice simulations
\cite{DamgaardHeller,EvertzJersakKanaya} for large $g$. The second order transition for $g=0$ is the $O(4)$-model transition 
\cite{TetradisWetterich}.}\vspace*{1cm}\\
It was established that for large $\lambda$ and large $g_4^2$ the phase transition observed for $T=0$ by varying $M_H$ turns to a crossover for (fixed) large $T$.
From this finding it was argued that the temperature dependent transition at fixed $M_H$ should also be a crossover for large $\lambda$ and $g_4^2$.
Direct access to the order of the high temperature phase transition for small\footnote{Realistic values of $g_4^2$ correspond to the ``deconfinement region'' for high $T$ in the language of
\cite{DamgaardHeller,EvertzJersakKanaya}.
The nonperturbative effects for the ``confinement region'' for large $g_4^2$ discussed in 
\cite{DamgaardHeller,EvertzJersakKanaya}
should not be confounded with the ``strong interaction effects'' arising even for small $g_4^2$ due to the flow of $g^2(k,T)$. In particular, for the ``confinement region'' for large $g_4^2$ the relevant scale is set by the $T=0$ confinement scale rather than by $k_s \sim \alpha_w T$ unless the temperature is far beyond the critical temperature.} $g_4^2$ - as realized in nature - requires, however, an understanding of the renormalization flow of couplings which was not available at this time
\cite{EvertzJersakKanaya}.  
The main idea for the proposal 
\cite{ReuterWetterichNP}
that crossover is relevant for the electroweak phase transition despite the small value of $\alpha_w$ relies on the result that even for small $\alpha_w$ the effective coupling $g^2(k,T)$ or $g_3^2(k)$ becomes strong if the effective infrared cutoff $k$ is small enough.
Near the transition line (hypersurface) in the phase diagram (fig.~2) the effective couplings then flow into the region of the phase diagram for which instead of a phase transition one has a crossover situation.
For a prediction of electroweak crossover based on the combination of the running of $g(k,T)$ with the crossover results from the lattice studies
\cite{DamgaardHeller,EvertzJersakKanaya}
for large $g_4$ it is important that $g(k,T)$ flows to large values on {\it{both}} sides of the transition line.
Only in this case the model with small $g_4$ can be effectively mapped into a model with large $g$ and lower short distance cutoff (larger lattice spacing), as simulated in 
\cite{DamgaardHeller,EvertzJersakKanaya}.
The details therefore depend on the value of $M_H$
as can be seen by comparing the running of $g^2(k,T)$ with $\lambda(k,T)$ as $k$ is lowered
\cite{WetterichSintra}.
For small $M_H$ (small initial $\lambda$ at $k = \pi T$) the quartic scalar coupling vanishes at some scale where $g^2$ is still small (cf.~fig.\,1 for $\overline{M}_H = 35$~GeV).
This scale becomes a characteristic scale for the first order phase transition which provides the effective infrared cutoff in the low temperature phase for $T=T_c$.
(This may be called a three dimensional Coleman-Weinberg
\cite{ColemanWeinberg}
effect.)
The gauge coupling stops its increase at this scale and is therefore not strong in both phases.
In the opposite case, for very large $M_H$, the gauge coupling becomes strong before the quartic scalar coupling vanishes.
The whole dynamics of the transition is then characterized by a strong gauge coupling and one expects crossover. 
There must be a critical value $M_H^{(c)}$ where the line of first order transitions (in the $M_H-T$-plane) ends such that for $M_H < M_H^{(c)}$ the transition is first order whereas for $M_H > M_H^{(c)}$ one has crossover.
At the endpoint $M_H^{(c)}$ the phase transition should be second order.
A naive estimate for $M_H^{(c)}$ could take the value where $\lambda(k,T)$ vanishes at the scale $k_s$ characteristic for the gauge coupling becoming strong.
From fig.~1 one sees that this happens for $\overline{M}_H \simeq 80$~GeV.

The crucial importance of the value of $M_H$ can also be understood
from the viewpoint of the effective three-dimensional theory.
In three dimensions, the effective scalar and gauge couplings
have dimension of mass, $\bar\lambda_3=\lambda_4T,
\bar g_3^2=g_4^2T$. The model is specified once the classical
action is given for a given
ultraviolet cutoff which we associate with $\pi T$,i.e.~$\bar\lambda_3, 
\bar g^2_3$, and the scalar mass
term $\bar m_3^2$ are fixed. We note that the ultraviolet cutoff is not
arbitrary here since the three-dimensional model which obtains
from dimensional reduction from high temperature quantum field
theory cannot be extended beyond momentum scales $ \sim \pi T$.
From the couplings we can form two dimensionless ratios, $x
=\bar{\lambda}_3/\bar g^2_3$ and $z=\bar g^2_3/\pi T$. If we select
a small range in $y=\bar m^2_3/\bar g_3^2$ which corresponds to
temperatures very near $T_c$ the characteristics of the phase
transition can still depend on $x$ and $z$. Let us first look
at the universality limit of infinite ultraviolet cutoff,
i.e. $z=0$. For this limit the phase diagram can only depend
on $x=\overline{M}^2_H/(4 M^2_W)$ and $y$. In particular, the existence of
a crossover region only depends on $\overline{M}_H^2/M_W^2$, but is independent
of the value of $\bar g^2_3$ which only sets the scale. If there exists
a crossover region for large $\bar g^2_3$, there should also
exist one for arbitrarily small $\bar g_3^2$! In the real world
$z$ is not zero, but determined by the electroweak fine
structure constant $z=4\alpha_w$. For small $\alpha_w$ the
characteristics of the transition can be described by
an expansion in $z$ which accounts for the nonuniversal effects.
An establishment of a crossover region in the phase diagram
in the universal limit $z=0$ is then a very strong argument
in favor of crossover  for small $z$\footnote{The early
lattice simulations 
\cite{DamgaardHeller,EvertzJersakKanaya}
found crossover
for high $z$ and this cannot be extra\-polated directly to $z\to 0$.}!

There is an analytic approach based on gap equations
\cite{BuchmuellerPhilipsen}
which seems to us very
promising for a quantitative description of the crossover
region as well as the endpoint of the first-order line
$M^{(c)}_H$ and $M_H$ somewhat below $M^{(c)}_H$. It determines the magnetic
mass of the $W$-bosons (as well as all other mass terms of the
model) by the solution of a gap equation
\bea
M_W^2(T)=M_W^2(0)+\Sigma(M_W^2(T))
\label{8}
\eea
with $M_W^2(0)$ the $W$-boson mass for $T=0$ and $\Sigma(M_W^2)$
computed in a resummed one-loop approximation, which involves
in turn the temperature-dependent mass terms. For $T$
near $T_c$ and $M_H$ smaller than some critical value
$M_H^{(c)}$ it was found 
\cite{BuchmuellerPhilipsen}
that the gap equation has two solutions, associated
with the high and low temperature phases, respectively.
This is the picture one expects for a first-order transition,
with one of the phases being metastable. For $M^{(c)}_H \simeq 80$~GeV
these two solutions merge into one as one would expect for
the endpoint of a line of first-order transitions beyond
which the transition becomes a crossover. Even though the
computation was performed in a particular gauge, thermodynamic
quantities and the value of $M_H^{(c)}$ are gauge-invariant
quantities and should not depend on the gauge. For the high
temperature phase the value of the magnetic mass was obtained
$M_W \sim g_4^2 T \sim k_s$, as it should be. 
(The proportionality coefficient came out too small, but this may be cured
by using a running gauge coupling $g(k,T)$ instead of $g_4$.)
We note that the
inclusion of higher-order or nonperturbative corrections
in $\Sigma$ may change $M^{(c)}_H$, but the overall picture
seems to be near to what one would expect. Furthermore, the
nonvanishing magnetic mass in the high temperature phase
has been associated with a nonperturbative expectation value
of the Higgs doublet even for $T>T_c$ 
\cite{BuchmuellerPhilipsen}. 
One is then left
with a picture where the two coexisting phases at $T_c$ correspond
to two distinct minima of the free energy as a function
of some doublet field. (This may be a combination of the
original Higgs doublet and a nonperturbative composite operator.)
Both of the minima correspond, however, to a nonvanishing
expectation value of this field - in contradistinction to
the traditional picture where the doublet expectation value
vanishes in the high temperature phase. An attempt for an
explanation of such a picture in terms of nonperturbative
condensates can be found in
\cite{BergerhoffWetterichParis}.

In summary, the analytical considerations provide strong
support for the idea of a crossover for large $M_H$.
Unfortunately, none of these arguments is quantitatively
very precise, and in particular the error on $M_H^{(c)}$ is
not known -- values of $M^{(c)}_H$ between 70 and 150~GeV seem
to be perfectly consistent with these approaches. It needed
numerical lattice simulations to provide a definite answer
to these questions. Recently, three-dimensional simulations
\cite{KajantieEtAlLargeMH}
have sett\-led the issue: The
important result is the determination of the critical ratio
$x_c=(M_H^{(c)}/(2M_W))^2$ for the three dimensional $SU(2)$ Yang-Mills-Higgs
system in the universal limit $(z=0)$. Within the high temperature
standard model, this corresponds to $M_H^{(c)}$ near $80$~GeV. 
In terms of the phase diagram in figure 3 this establishes that for $M_H > M_H^{(c)}$ the transition is a crossover in the immediate vicinity\footnote{The correlation length of the high temperature field theory becomes nevertheless large in this limit, being proportional $(g^2 T)^{-1}$.} of $g=0$, i.e. for $g \rightarrow 0^+$.
This
result has been confirmed by a different simulation for gauge-fixed
observables
\cite{KarschEtAl}
and by using refined criteria for the determination of the crossover point
\cite{GuertlerIlgenfritz}.
The central value of $M_H^{(c)}$ is found 
\cite{GuertlerIlgenfritz}
as $72.2$~GeV for the pure $SU(2)$-Higgs model.
In the standard model with $m_t = 175$~GeV this corresponds to $M_H^{(c)} = 72$~GeV, with a critical temperature $T_c = 110$~GeV.
In view of the present experimental lower bound $M_H > 70$~GeV
\cite{Jerusalem}
one concludes that for the standard model the electroweak transition in the early universe is a crossover or very near to a crossover!
The main reason for the
feasibility of these lattice studies is the very existence
of the crossover itself: For $\overline{M}_H$ sufficiently large compared
to $M_H^{(c)}$ all correlation lengths are small enough to fit into
the volume of the lattice. In that respect the situation is
similar to simulations for $\overline{M}_H$ much smaller than $M_H^{(c)}$.
Only the vicinity
of $M^{(c)}_H$ is difficult to access directly, since for the endpoint
the correlation length diverges. Having established the
character of the transition both for $\overline{M}_H<M_H^{(c)}$ (first order)
and $\overline{M}_H>M_H^{(c)}$ (crossover), the situation is nevertheless
unambiguous.

After the establishment of crossover as the most salient
prediction of the picture with strong interactions at the
electroweak phase transition, one may wonder if other
features of this scenario can also be verified by lattice
simulations. The first is the suggestion
\cite{WetterichSintra}
that the
properties in the symmetric phase are almost independent
of the value of $M_H$, being determined dominantly by the scale
$k_s$. In particular, the masses of the $W$-boson excitations
should scale $\sim \alpha T$.
This seems indeed
to be confirmed by the simulations:
The $W$-boson mass in the high temperature phase is
by now determined consistently by several groups 
\cite{KarschEtAl,Owe,MassesInSymmetricPhase}
and turns out to be around $2k_s$ if $k_s$ is given by
(\ref{5}). This may be taken as a confirmation that the computation of
the ``confinement scale'' is roughly correct. 
Actually, since we did not give any precise definition of $k_s$, we may use the $W$-boson mass in the limit $T\rightarrow \infty$ as the physical scale replacing $k_s$.

Another important
prediction of the ``strongly interacting electroweak phase
transition'' is the rise of the effective gauge coupling as
the infrared scale $k$ is lowered. One may test this in the low
temperature phase where $k$ is given by $M_W$ and depends on $T$
and $M_H$. Unfortunately, it is not easy to measure the three-gauge
boson vertex directly on the lattice. Instead, one can use the
observation that the ratio $2M_W^2(T)/[Z_\varphi(M_W,T)\rho_0
(T)]$ (eq. (\ref{6})) can be used as an alternative definition of a
renormalized gauge coupling $g^2(M_W,T)$. The value of $M_W(T)$ is
accessible to lattice simulations. We may define $(M_W^{(tree)})^2
=\frac{1}{2}g_4^2\rho_0(T)$ and consider the ratio $M_W(T)/M_W
^{(tree)}(T)=(g(M_W,T)/g_4)Z_\varphi^{1/2}(M_W,T)$. We have
computed this ratio for various values of $\overline{M}_H$ at the critical
temperature with a running gauge coupling according to (\ref{4}).
The value of $\tau=1.6$ has been chosen to match the lattice
value for $\overline{M}_H=180$~GeV. 
We also have taken $g(\pi T,T) = g_4$ and $Z_\varphi(\pi T,T) = 1$ which holds only up to corrections $\propto g^3$.
A comparison with lattice data
\cite{KajantieEtAlLargeMH,MassesInSymmetricPhase,LainePC}
is shown in figure 4, with data points at $\overline{M}_H = 180,\,120$ and $60$~GeV.
Here the three
curves reflect the dependence of $Z_\varphi$ on $\overline{M}_H$, with
the upper curve for $\overline{M}_H=180$~GeV and the lower one for
$\overline{M}_H=60$~GeV. In view of the uncertainties, the agreement
is satisfactory. A more sensitive test would become
available if the flow equation for $g^2$ and the initial value for $k=\pi T$ are computed for
the particular definition (6), i.e. if $\tau$ is determined
by an analytical calculation.
\begin{figure}
\hspace*{1.3cm}\epsfig{file=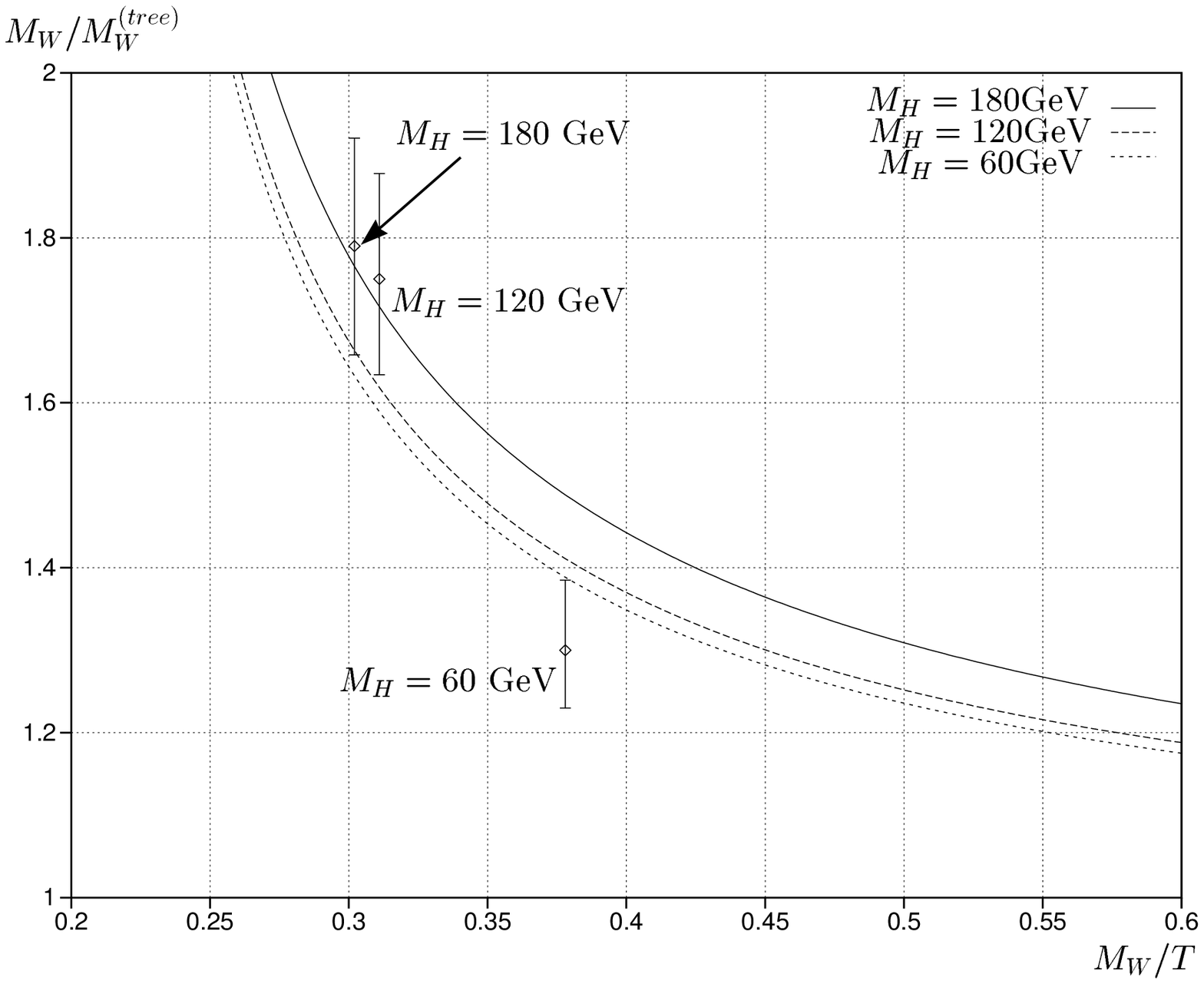,width=11.5cm}
\myfigcaption{The running gauge coupling defined through the ration $M_W/M_W^{(tree)}$ in comparison with lattice results 
\cite{KajantieEtAlLargeMH,MassesInSymmetricPhase,LainePC}.}
\end{figure}
\afterpage{\clearpage{
\begin{figure}[H]
\begin{center}
\epsfig{file=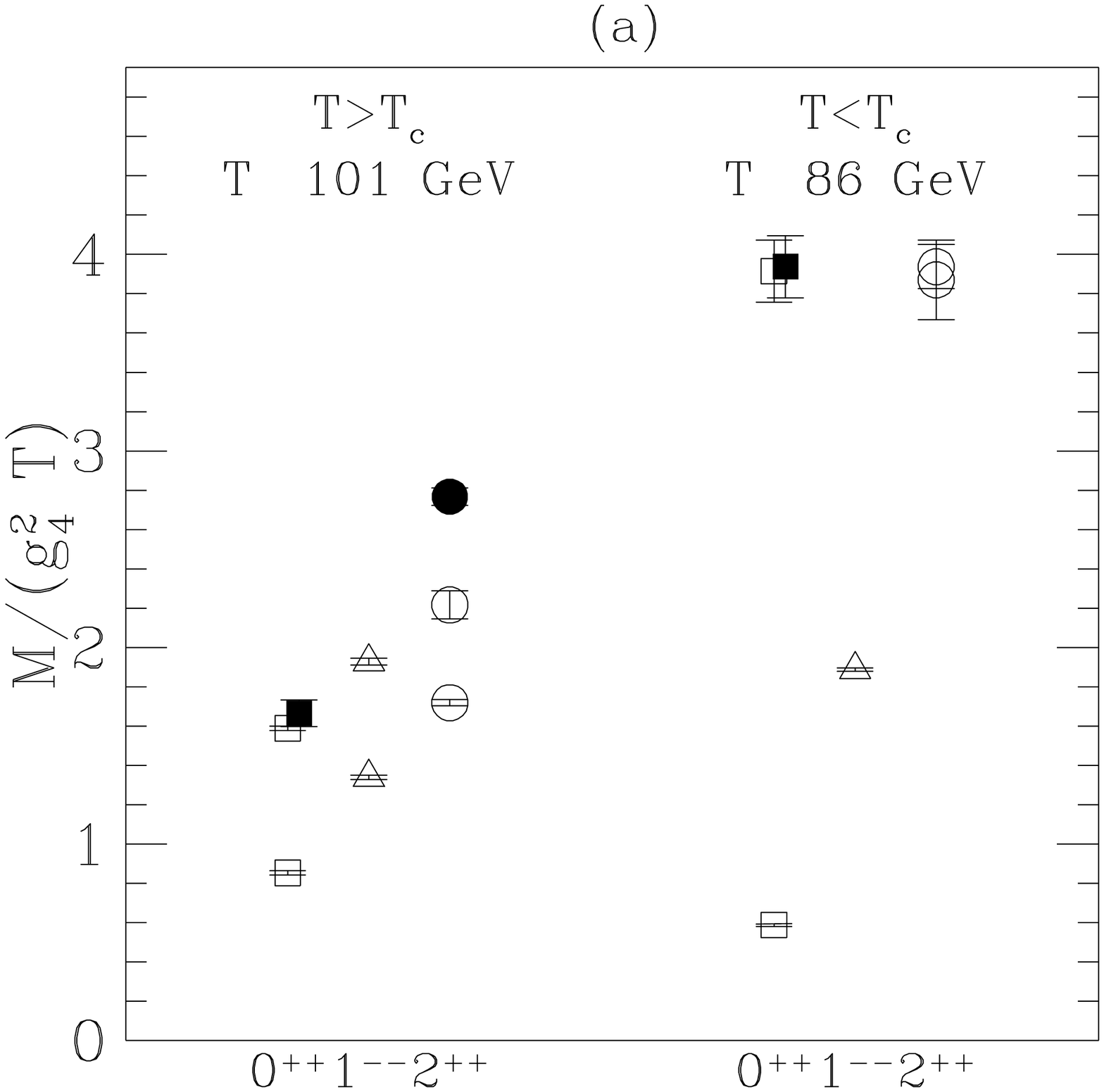,width=8cm}
\put(-177,192){$\scriptstyle{\simeq}$}
\put(-78.5,192){$\scriptstyle{\simeq}$}\\
\epsfig{file=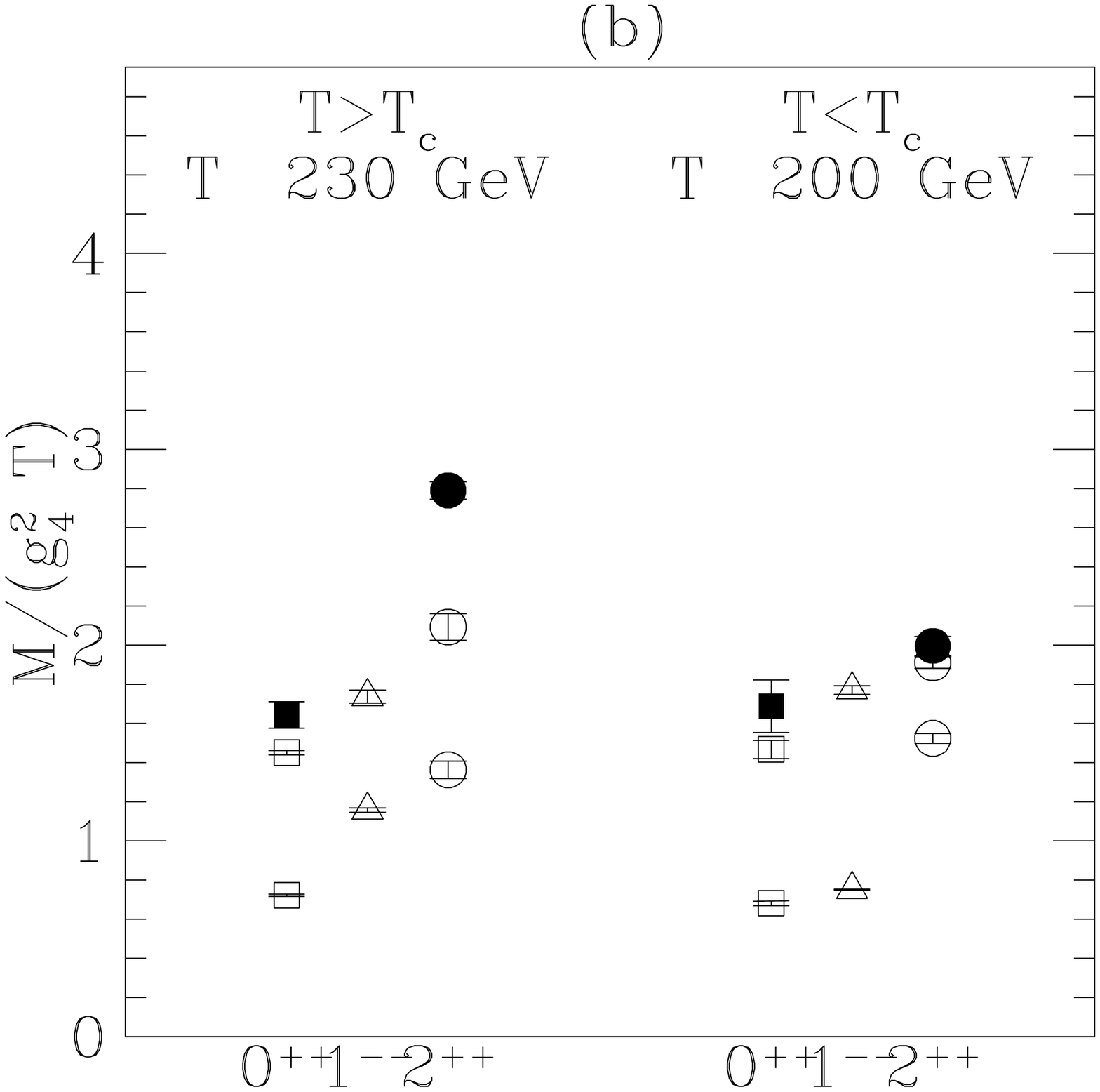,width=8cm}
\put(-179.5,192){$\scriptstyle{\simeq}$}
\put(-85,192){$\scriptstyle{\simeq}$}
\end{center}
\myfigcaption{Spectrum of scalar ($0^{++}$), vector ($1^{--}$) and tensor ($2^{++}$) excitations
(from \cite{Owe}).
The lowest $0^{++}$ and $1^{--}$ state may be associated with the scalar and W-boson.
Full symbols denote states in the pure $SU(2)$ gauge theory (``W-balls'').
Fig.~5a corresponds to a low scalar mass $\overline{M}_H = 35~\mathrm{GeV}< M_H^{(c)}$~($\lambda_3 / g_3^2 = 0.0239$) where $T_c = 93$~GeV.
Fig.~5b obtains for a large Higgs mass $\overline{M}_H = 120~\mathrm{GeV}>M_H^{(c)}$~($\lambda_3 / g_3^2 = 0.274$), with $T_c = 213$~GeV.
The qualitative difference between the high and low temperature spectrum has disappeared.}
\end{figure}}}
Nevertheless, there seems to
be a lattice confirmation for a running effective gauge coupling
already at the present stage. We conclude that the
picture of strong interactions at the electroweak phase transition
seems to be consistent with the presently available lattice
results.
We observe that for large $\overline{M}_H$ the value of $g(M_W,T)$ settles at some value $g_{HT} \simeq 1.8 g_4 \simeq 1.2$.
Since for $\overline{M}_H > M_H^{(c)}$ we may look at the situation at $T_c$ also from the viewpoint of the high temperature phase, $g_{HT}$ can be associated with the value of the gauge coupling for which a dynamical magnetic $W$-boson mass is generated.
The reader may be surprised that $g_{HT}$ turns out to be relatively small, but it should be remembered that the relevant parameter for perturbation theory is $g_3^2(M_W) = g_{HT}^2 T / M_W(T) \simeq 4.9$.

Finally, the strong interaction picture predicts
\cite{ReuterWetterichNP,WetterichSintra}
for the high temperature phase a rich spectrum of excitations in analogy to pure QCD, containing W-balls etc..
These excitations should be absent in the low temperature phase for scalar masses sufficiently below $M_H^{(c)}$ where the gauge coupling remains small.
On the other hand, for a scalar mass near or above $M_H^{(c)}$ the spectrum in the low temperature phase should be very similar to the one in the high temperature phase
\cite{WetterichSintra}.
Furthermore, the ratios mass/T for the various excitations are expected to depend only weakly on $T$
\cite{WetterichSintra}.
All these features are fully confirmed by lattice simulations.
Results of a recent high accuracy simulation 
\cite{Owe}
are reproduced in figure 5.
For the lower vector ($1^{--}$) state the ratio $M_w/T$ is almost independent of $\overline{M}_H$ and $T$ in the high temperature phase.
The existence of scalar and tensor W-balls with properties similar to the pure $SU(2)$ gauge theory is clearly established.
Furthermore, for $\overline{M}_H > M_H^{(c)}$ the spectrum on both sides of the (pseudo-)critical temperature is quite similar.

Having learned a great deal about the detailed behaviour of the
temperature dependence of the parameters of the standard model and
a possible electroweak phase transition, let us finally come
back to the basic question which started all these
investigations: Is there really a symmetry restauration\footnote{Note that symmetry restauration in a generalized sense is, in principle, possible even for gauge theories where no gauge invariant order parameter exists in the standard sense. An example is presumably the abelian Higgs model.} of the
electroweak  $SU(2)$-symmetry at high
temperature, as originally proposed by Kirzhnits and Linde
\cite{KirzhnitsLinde}? 
The observation of the phenomenon of crossover for
large $M_H$ suggests that the answer may be negative! For a
possible observation of symmetry restauration we should include
the photon with a nonzero gauge coupling $g'$ of the gauge
boson corresponding to hypercharge. For low $T$ there is
a mass split between $M_W$ and $M_Z$ given by
the Weinberg angle $\sin\vartheta_W$. Symmetry restauration would
mean that for high $T$ the masses of $M_W$ and $M_Z$ become
degenerate. We also can study the mass split between
top and bottom quarks, which should vanish at high $T$ in
case of symmetry restauration. Crossover gives a different
picture. We propose that crossover persists in presence of the hypercharge
gauge boson -- this needs to be verified but we would not
expect\footnote{The correlation length is finite for $M_H \gg M_H^{(c)}$ and $g'=0$.
Turning on $g'>0$ continuously, the correlation length must remain finite at least for a certain range of $g'$.
There is no reason to expect a new type of transition associated with the hypercharge sector.
For $g'>0$ one simply has an explicit breaking of global $SU(2)_R$-symmetry.}
that a small $g'$ changes the dynamics of the transition.
We may now approach $T_c$ from below near the critical scalar
mass $M_H^{(c)}$. There is no reason why the top-bottom mass split
should vanish for $T\to T_c$. If this is true, however,
analyticity implies a nonvanishing mass split also for
$T>T_c$! We are back to the picture of a nonvanishing
expectation value of a scalar doublet operator also in the
high temperature phase
\cite{BuchmuellerPhilipsen}. 
By dimensional reasons, this expectation
value can only be proportional to $T$ in the limit $T \rightarrow \infty$. 
One concludes that the
bottom-top mass split increases for large temperature
$\sim T$ and $SU(2)$ symmetry is never restored! The situation for
the $W$- and $Z$ bosons is a bit more subtle: Again,
one expects that a mass split remains for arbitrarily
high $T$. It is, however, given by a temperature-dependent
Weinberg angle $\sin\vartheta_W(T)$ which may be
considerably smaller at large $T$ as compared to $T=0$, due to the
increase in $g(T)$ and the decrease of the abelian coupling
$g'(T)$
\cite{ReuterWetterichNP}.
\setcounter{footnote}{0}

If one accepts that $SU(2)$-symmetry is
not restored and there remains a nonvanishing doublet
expectation value at high $T$, one arrives at a
semiquantitative picture. Let us denote by $\rho_{LT}$
the square of the renormalized doublet expectation value\footnote{This is not a gauge invariant quantity if computed with a particular gauge fixing. There is, however, a gauge invariant counterpart to it.}
in the low temperature phase $(\rho_{LT}=\rho_0(T)
Z_\varphi(M_W,T))$ and similarly by $\rho_{HT}$ the one
for the high temperature phase. Particle masses and
splittings are then given by temperature-dependent effective
couplings in the low- and high-temperature phases, respectively,
\bea
&&M^2_{W,LT}=\frac{1}{2}g^2(M_{W,LT},T)\rho_{LT}\nonumber\\
&&M^2_{W,HT}=\frac{1}{2}g^2(M_{W,HT},T)\rho_{HT}\nonumber\\
&&(m_t-m_b)_{LT}=(h_{t,LT}-h_{b,LT})\sqrt{\rho_{LT}}\nonumber\\
&&(m_t-m_b)_{HT}=(h_{t,HT}-h_{b,HT})\sqrt{\rho_{HT}}
\label{9}
\eea
We do not expect that the Yukawa couplings in the low and high
temperature phases differ substantially. We also remind
$M_{W,HT}\sim k_s\sim g_4^2T$ and that $g_{HT}=g(M_{W,HT},T)$
is a universal number independent of $T$ in the limit $T\rightarrow \infty$. 
We propose here that $g_{HT}$ is actually independent of $M_H/T$ and $\frac{T-T_c}{T}$ in a good approximation.
If $g_{HT}$ can
be computed by the solution of a flow equation (cf.~eq.~(\ref{4}) and
fig.~4) one can fix $\rho_{HT}\sim(g^4_4/g^2_{HT})T^2$ and
make a prediction for the top-bottom mass split,
$m_t-m_b\sim((h_t-h_b)g^2_4/g_{HT})T$. Since $g_4/g_{HT}<1$,
$(h_t-h_b) g_4<1$ one concludes that this mass split is
substantially smaller than  the effective mass $\sim \pi T$
for the lowest excitation. Even though $SU(2)$-symmetry
is never really restored at high $T$, the actual situation
is not very far from degenerate masses in
the multiplet! Also for low enough $M_H<M_H^{(c)}$ the first-order
phase transition looks quite close to the picture of symmetry
restauration: At $T_c$ the ratio
$(m_t-m_b)_{HT}/(m_t-m_b)_{LT}\sim(\rho_{HT}/\rho_{LT})^{1/2}
\sim g_4^2T_c/(g_{HT}\sqrt{\rho_0(T_c)})$ is much smaller than
one
(cf.~the table with $\sqrt{2 \rho_0(T_c)} \sim v$ for low enough $\overline{M}_H$)! 
Of course, as $M_H$ increases, $\rho_{LT}$ decreases
and $\rho_{HT}$ increases until they are equal at $M^{(c)}_H$.

One may actually turn the relations (\ref{9}) into relations
between the $W$-boson masses in the high and low temperature
phases, testing in this way the simple strong interaction
picture with a universal constant $g_{HT}$. 
For $M_H=M_H^{(c)}$ the values of $\varphi_{HT}$ and $\varphi_{LT}$
should coincide for the critical temperature and one finds
\cite{BergerhoffWetterichErice}
\bea
\varphi_{LT}= \varphi_{HT} \simeq 0.55 T_c
\label{12}
\eea
The $W$-boson mass in the high temperature phase and, for $M_H > M_H^{(c)}$, also in the low temperature phase has been estimated along these lines 
\cite{BergerhoffWetterichErice}
\bea
M_{W,HT} = a g_4^2 T
\label{new12}
\eea
with $a \simeq 0.8-0.9$.
This compares well with the lattice value $a \simeq 1.1$ for the high temperature phase and $M_H < M_H^{(c)}$ and $a \simeq 0.9$ for the low temperature phase and $M_H > M_H^{(c)}$ (see fig.~5)!

We may summarize our conclusions in the following points:

(1) The standard model in thermodynamic equilibrium is by now well
understood (at least for $\sin\vartheta_W=0$). This was made
possible by a combination of analytical and numerical efforts. In
particular, dimensional reduction to an effective three-dimensional
theory combined with numerical lattice simulations in three
dimensions has provided results with high quantitative accuracy
for the most important  thermodynamic quantities.

(2) For a mass of the Higgs scalar $M_H<M_H^{(c)}\simeq 70$~GeV there
is a first-order phase transition to a qualitatively different
high temperature phase of electroweak interactions. This phase
transition disappears for $M_H>M_H^{(c)}$. The rapid qualitative
change is now described by an analytic crossover, very
similar to the vapor-water transition beyond the
critical pressure.

(3) There is presumably no true symmetry restoration of the electroweak
$SU(2)$-symmetry in the high temperature phase. Nevertheless, different
members of $SU(2)$ multiplets have approximately degenerate masses for
high $T$.

(4) The baryon asymmetry cannot be generated during the electroweak
phase transition in the standard model. There is no deviation
from thermodynamic equilibrium for $M_H>M_H^{(c)}$.
Possible non-equilibrium effects for $M_H < M_H^{(c)}$ are by far too weak for a realistic scalar mass $M_H > 70$~GeV.
This result is very important, since it indicates the necessity
of an extension of the standard model! There are two general
possibilities: Either the scalar sector is extended within
the model valid below 1 TeV, as for example in
supersymmetric theories. For appropriate regions in SUSY-parameter
space 
\cite{SUSY}
the electroweak phase transition
can be strongly first order such that baryogenesis may occur in
bubble walls if also the CP-violation is sufficiently strong. As an alternative,
there may be an extension of the standard model which generates an
asymmetry in $B-L$, as for example in certain grand unified
theories.

(5) The physics in the high temperature phase is characterized
by effective strong interactions, despite the small values
of the zero temperature dimensionless couplings. This also holds
for the low temperature phase near $T_c$ if $M_H$ is in the vicinity
of $M_H^{(c)}$. These strong interactions are compatible with a partial
convergence of perturbation theory in low order.

(6) The characteristic scale in the high temperature phase is
set by a three-dimensional ``strong interaction scale'' 
$k_s\simeq g_4^2T$. In principle, thermodynamic quantities and masses of the
excitations could also depend on the additional parameters
$\frac{T-T_c}{T}$ and $M_H/T$. Except for the characteristics of scalar
excitations a simplified picture assumes that the dependence
on these additional parameters can be neglected. The
high temperature phase resembles then the one in a Yang-Mills
theory without scalars. This picture provides a satisfactory
understanding of several results from lattice simulations.
It predicts, in particular, a value for the $W$-boson mass in the high temperature phase which agrees well with the lattice results.

\bigskip

\noindent {\bf{Acknowledgement}}

\noindent We would like to thank O. Philipsen for making available to us figure 5.

\end{document}